\documentclass{icrc2009}

\usepackage{graphicx}   
\usepackage[caption=false]{caption}    
\usepackage[font=footnotesize]{subfig} 
\usepackage{fixltx2e}
\usepackage{url}

\newcommand{\shorttitle}[1]%
{\markboth{Proceedings of the 31\MakeLowercase{$^{st}$} ICRC, {\L}\'{o}d\'{z} 2009}{#1} }
\newcommand{\etal}{\MakeLowercase{\textit{et al. }}} 

\begin{document} \title{A conceptual design of an advanced 23 m diameter IACT
of 50 tons for ground-based gamma-ray astronomy}

\author{\IEEEauthorblockN{Eckart Lorenz\IEEEauthorrefmark{1,2}, Daniel
Ferenc\IEEEauthorrefmark{3}, M. Victoria Fonseca\IEEEauthorrefmark{4}, and Robert Wagner\IEEEauthorrefmark{1}} \\
\IEEEauthorblockA{\IEEEauthorrefmark{1}Max-Planck-Institut f\"ur Physik,
F\"{o}hringer Ring 6, D-80805 Munich, Germany}
\IEEEauthorblockA{\IEEEauthorrefmark{2} ETH Z\"urich, CH-8093 Zurich,
Switzerland}
\IEEEauthorblockA{\IEEEauthorrefmark{3}University of California,
Davis, One Shields Ave., Davis, CA-95616, USA}
\IEEEauthorblockA{\IEEEauthorrefmark{4}Universidad Complutense, Av. Complutense
s/n, E-28040 Madrid, Spain}}

\shorttitle{Lorenz \etal Design of a 23 m IACT}
\maketitle

\begin{abstract}
 A conceptual design of an advanced IACT with a 23 m 
diameter mirror and of 50 tons weight will be presented. A system photon 
detection efficiency of 15-17{\%}, averaged over 300-600 nm, is aimed to 
lower the threshold to 10-20 GeV. Prospects for a second generation camera 
with Geigermode apds will be discussed.
\end{abstract}


\section{Introduction} 

The window of very high energy (VHE) ground-based gamma-ray astronomy was 
opened 1989 by the Whipple collaboration observing TeV gamma-rays from the 
CRAB nebula \cite{weekes}. In the following 20 years this research field
developed  with a breath-taking speed, and up to now more than 70 sources have
been  discovered and many new and fundamental physics results attained. The 
success was mainly achieved by improving the so-called imaging atmospheric 
Cherenkov telescopes (IACT), i.e. by making it possible to detect Cherenkov
light  from extended air showers generated by cosmic particles in the upper 
atmosphere. One of the clear messages from the current results is that  further
progress can be made if detectors are further improved. The main goals are
lowering the energy threshold from currently about 50-60 GeV down  to close to
10-20 GeV, and to increase the sensitivity by a significant  factor compared to
the currently best instruments. The sensitivity can be  raised by two methods:
a) by increasing the photoelectron detection rate  from showers, thus improving
mainly the gamma/hadron separation, and b) by  operating many IACTs in an array
configuration, i.e. by stereo observation  along with a significant increase of
the collection area. Another goal is to  achieve a significant overlap in
energy with the FERMI satellite \cite{nasa}. Here  we present a conceptual
design for a next generation IACT with improved  performance, which could be
the basic element for the above mentioned goals.  Additional aims are the rapid
positioning which will permit  observations of GRBs in near  real time after a
satellite alert, and extending the observation of strong  sources in the
presence of moonlight. Using such a telescope in an array  configuration should
allow one to achieve a significantly higher  sensitivity. In the following, we
will mention only some of the currently  most important questions in the area
of fundamental physics studies and  searches for new gamma-ray sources:

\begin{itemize}
\item Dark matter searches 
\item Quantum gravity studies, i.e. searches for possible Lorentz invariance violation
\item Study of the extragalactic background light (EBL)
\item Searches for topological defects 
\item Study of the upper energy range of GRBs
\item Study of the upper energy range of pulsars
\item Search and studies of high red-shift AGNs
\item Studies of flaring AGNs and AGNs in their `low' state of emission
\end{itemize}

Obviously, the program will be much richer because of the increased  discovery
potential for new sources due to improved sensitivity, but the scope of
this contribution has to be limited to the above mentioned topics. Although
there is partial  overlap with FERMI, most of the studies require sensitivities
above $\approx$  30-50 GeV to be orders of magnitude higher than what can be
achieved with  FERMI.

\section{The basic telescope concept}

The basic design is a derivative of the MAGIC telescope \cite{baix}, but with a much 
larger mirror area, lower weight and a number of technical improvements. Fig. 
1 shows a conceptual drawing of the alt-azimuth telescope with a 23 m 
diameter parabolic mirror and an f/D of 1.2. The weight of the moving part 
of the telescope will be only 50 tons in order to allow rapid rotation. Such 
a low weight can be achieved by using for most of the mount carbon fiber 
reinforced plastic (CFRP) tubes, low weight mirror panels of $\approx $ 2 
m$^{2}$ area and compact undercarriages running on an I-beam serving also as 
an anchorage against liftoff of the telescope during strong storms. The 
telescope will follow the `semisoft' design of MAGIC, i.e. correct small 
deformations of the structure during source tracking and caused by temperature 
changes by using an improved active mirror control and also an active camera 
support mast control (see below).

\subsection{The lower mount and the mirror support structure}

The telescope is a so-called alt-azimuth design running on 6  undercarriages on
a circular I-beam fixed with a number of steel bars to  the central axis. The
azimuth structure resembles very much the  configuration of MAGIC but uses
large diameter high strength CFRP tubes,  which can now be produced in series
by industry. The bogeys will be very  compact and a factor 3 lighter compared
to the MAGIC ones. For the main  mirror support dish a modified configuration
of the tubular construction  will be used. The formerly used basic element of a
pyramid shape will be  replaced by a significantly stiffer tetraeder
configuration, thus allowing a  stiffer spaceframe for the 23 m diameter
support, but using the same length  of tubes compared to the MAGIC
construction. The spaceframe will be a  4-layer configuration. The CFRP
material is considerably more expensive than steel, but a large weight
saving, lower transport and installation  costs and much lower drive costs,
eventually will balance the steel  construction costs. CFRP has, besides
a much lower weight, two other  advantages: a) the thermal expansion is nearly
zero; thus dish deformations  due to temperature changes are small, and b) the
material has a much higher  oscillation damping, thus permitting the use of
longer tubes of a smaller diameter.  In total, the construction will use about
2000 m tubes of 80 or 100 mm  diameter for the space frame and about 600 m of
250 mm diameter high  strength tubes ($>$1000 tons breaking resistance when
pulling and $>$130  tons buckling strength) for the azimuth structure. A drive
system similar to  that of MAGIC will be sufficient because of the lower weight
of only 50 tons  compared to MAGIC's 70 tons.

 \begin{figure}[!t]
  \includegraphics[width=2.5in]{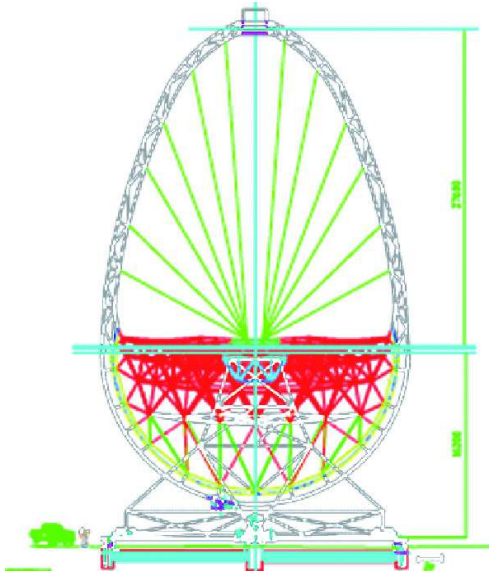}
  \caption{Layout of the 23 m telescope (P. Sawallisch)}
  \label{fig1}
 \end{figure}

\subsection{The mirror and active mirror control}

The 420 m$^{2}$ mirror will be composed of 210 panels of 2 m$^{2}$ each.  The
gross mirror profile will be parabolic to minimize any additional time  spread
of the Cherenkov light flash. Because of their size the mirror panels  have to
be aspheric. Two options for the mirror production are under  discussion: a)
use of diamond turned panels which can now be fabricated with  the shape of an
off-axis paraboloid, b) the replica technique using cold  slumped glass-aluminum Hexcell core sandwiches. By using aspheric negatives  it is also
possible to make the replicas according to the required shape on  the mirror
dish position. The mirror panels will have a hexagonal shape to  fit the
tetraeder elements of the space frame. A panel weight of $<$  20kg/m$^{2}$ is
aimed for. For further details see \cite{doro}. As mentioned, the  semisoft
design concept requires active corrections of the panel  orientations. It is
planned to have a permanent control. Two options for monitoring the
orientation are under evaluation: a) infrared laser beams  mounted on the mirror
panels and pointing to diffuse reflectors mounted  asides the camera but not
seen by the IR insensitive camera photosensors;  the laser spots will be
consecutively observed by an IR sensitive CCD camera  and deviations from the
nominal positions determined and corrected  accordingly, and b) small CCD cameras
on all mirror panels will observe  reference LEDs (mounted in the focal plane
outside the camera) and determine  the deviation of the reference orientation
and initiate corrections. The  latter solution is more complex and more
expensive but allows for higher  flexibility.

\subsection{Camera and camera support mast}

It is planned to use a camera of 4\r{ } field of view (FOV), subdivided  into
pixels of 0.07\r{ } diameter (hex entrance pupil), i.e., comprising  around
2600 pixels. The camera will be placed in the focal plane at 27.5 m  distance
from the dish. This poses a major challenge for the construction of  the camera
support mast. As for MAGIC, it is intended to use a single arc  support mast of
only 2-3 {\%} shadowing of the main mirror. Because of the  large distance in f
and the heavy camera the mast will have a back-to-back  structure of two tubes
spaced by 80 cm. In order to avoid transverse  displacements of the mast, as
well as oscillations, the masts will be fixed  in place by a number of high
strength ropes made from either Dyneema (about  15 times stronger than steel
for the same weight, need UV protection) or  Carbon fiber rods (used in the
deep sea oil industry, about 5 times stronger  than steel for the same weight).
In order to correct for mast bending at  large zenith angles we will use an
active bending correction by pulling the  cables with actuators for the
necessary mast correction.

The baseline pixel units will comprise a hemispherical 1'' photomultiplier 
with a `super-bialkali' photocathode and a light catcher lined with a 
dielectric reflector foil of $>$ 96{\%} averaged reflectivity between 290  and
700 nm. For low aging and operation during partial moon light the PMT  will be
operated at low gains of 2-3.10$^{4}$ followed by a high bandwidth,  low noise
preamplifier of at least 20-26 dB gain. The PMTs will be treated  to maximize
the photon detection efficiency (PDE) by a) optimizing light  collection by the
PMT front window using either a frosted glass or a diffuse  lacquer, and b) by
maximizing the voltage between photocathode and first  dynode to increase the
photoelectron collection efficiency and to minimize  time spread in the
front-end electron optics. The camera will be sealed by a  UV transparent front
window of very low reflectivity (broadband  antireflective coating or
plasma-etched surfaces). It is estimated that the  combined use of the above
noted changes will improve the  light-to-photoelectron conversion by about
25-38{\%} compared to a standard  pixel configuration of frequently used flat
window bialkali PMTS. An overall  PDE of 15-17 {\%} of the incident Cherenkov
light between 300 and 600 nm  should be achieved. Besides the baseline
configuration quite a few other  photosensor options are under evaluation; see
also below about a second  generation camera with G-apds. A possible option
might be the use of multipixel  flat panel photosensors like the ARCALUX
proposal because of the potential  of lower prices \cite{felo}. A critical
issue is the question of how much or if all the  DAQ and trigger electronics
will be placed also in the camera. On the one hand,  one will have a very
compact arrangement, while on the other hand, one has to  pay the heavy penalty
of a substantially heavier camera and serious cooling  problems. Also servicing
and mid-life updates are difficult. Increasing the  weight of the camera will
require a much more reinforced and heavier support  structure, which, in turn,
will require a significantly heavier mount  construction, thus missing by a
sizeable margin the 50 ton weight goal. An  alternative might be converting the
PMT signals back to light by VCSELs and  rout the optical analog signals by
optical fibers to an easy accessible  place at ground. This worked very well
for MAGIC, where it was possible to  change the DAQ and to add a new trigger
logic already after 4 years of  operation thanks to the rapid progress in
electronics and computer power.  (It should be noted that from the time of
constructing MAGIC until today the  computer power and the data storage
capacity have improved by a factor of at  least 50). The two options are still
under discussion, and a  careful evaluation has to be made of the extra costs
for the optical fiber system and the cost saved by using the considerably
lighter and simpler mechanics, the reduced  cooling and the easier access and
servicing. The new generation VCSELs have a  much lower noise and less mode
hopping compared to the ones used in MAGIC.  Tests have shown that the VCSEL
signal can be stabilized over a temperature  range of 70\r{ }C. All optical
fibers can be bundled in 26 cables, each of a  thickness of a RG 213 coax
cable.

\subsection{The trigger and the DAQ}

It is expected that the night sky light background will generate a single 
photoelectron rate of 150-200 MHz per pixel when viewing a sky section  outside
the galactic plane. Observing objects on the galactic plane or  during partial
moon light will result in significantly higher rates.  Therefore, an efficient
trigger system is needed to select possible signals  from air showers. In the
past, the readout speed of the DAQ and the control  computer were demanding a
high trigger selectivity to reduce the rate to  below 1 kHz. Assuming that
Moore's law still holds in the coming years one  can expect another factor of
50 increases in computer power and a similar  drop in the price of storage
elements. Concequently, one can envision a rather  simple trigger logic
resulting in a rather high rate. An easy solution for  the readout would be the
use of many processors reading out only small camera  sections and storing the
data in large raid discs. An additional computer  farm can process the raw
events and reject quickly the large number of  accidental triggers while
selecting a few specific patterns of event  candidates. This event selection
might continue over daytime to allow for a  slower processing than the initial
trigger rate. In summary, the progress in  computer power would allow to roll
over complexity of costly hardware  trigger logic to a much more powerful and
flexible software trigger logic.  The hardware trigger logic could be similar
to the SUM trigger used by MAGIC  for the CRAB pulsar study \cite{aliu}. To
suppress accidental triggers from large  amplitude afterpulses, clipping before
summing can be used \cite{lorenz1}. As current  PMTs have a pulse width equal
or below the time structure of low energy  showers of $\approx $ 2-3 nsec, a
digitizing system of at least 1 GHz  sampling rate is needed if one wants to
use timing and narrow signal windows  for the NSB reduction. Such systems are
now available in the form of  switched capacitor arrays. An example is the DRS
4 chip developed at the PSI  \cite{ritt}. This chip with 1024 capacitors can be
clocked up to 6 GHz and can  record signals of 1-3 nsec width with a dynamic
range of $>$ 66 dB while  having a power consumption of $<$ 100 mW/channel.

\subsection{Some rough performance estimates}

In the absence of fully-fledged MC simulations one can extrapolate some 
performance from smaller size IACTs. A threshold of 20-30 GeV is estimated  for
the normal source search and $<$ 15 GeV for pulsar studies. The  sensitivity of
a single 23 m IACT possessing the above parameters should be below 1{\%}  of
the Crab flux for a 5 $\sigma $ signal and 50 h observing time, while an  array
of 9 should reach a sensitivity of 0.5-2 milliCrab.

\subsection{Price and construction time}

The price is very much a function of the number of telescopes. While a  single
telescope might cost around 8 M Euro, the price per unit for a small series 
production should be 5-6 M Euro,  with about 1/2 of such amount being spent on
the mount and  mirrors, and the other half on the camera, trigger and readout.
Assuming early prototyping the  construction of a single telescope should take
around 2 --3 years, while a  small series production of e.g. 9 units would
require twice the time.

\section{A second-generation camera}

In the long run it might be possible to replace the PMTs by  solid-state
Geigermode-avalanche photodiodes (G-apd). These very compact  photosensors have
the potential to double the PDE when folded by the  Cherenkov spectrum and the
telescope's optical parameters, i.e. it should be  possible to lower the
threshold to close to 10-15 GeV. First tests of  detecting Cherenkov light
flashes gave encouraging results \cite{biland}. G-apds need  only a low
operation voltage ($<$100 V), no magnetic shielding, show no  aging and can be
exposed fully biased to daylight without destruction. It is  estimated that 2-3
more years of development will be needed to reach a  high maturity level.
Current weaknesses are the small area, a high price,  and some operation
deficits in optical cross-talk, drift of the PDE as a  function of temperature
and bias voltage, and insufficient UV sensitivity  below 400 nm, see
\cite{{renk},{lorenz2}}.

\end{document}